# Physical portrayal of computational complexity

ARTO ANNILA[1,2,3,*]

[1]Department of Physics, [2]Institute of Biotechnology and [3]Department of Biosciences, FI-00014 University of Helsinki, Finland

Computational complexity is examined using the principle of increasing entropy. To consider computation as a physical process from an initial instance to the final acceptance is motivated because many natural processes have been recognized to complete in non-polynomial time ($\mathcal{NP}$). The irreversible process with three or more degrees of freedom is found intractable because, in terms of physics, flows of energy are inseparable from their driving forces. In computational terms, when solving problems in the class $\mathcal{NP}$, decisions will affect subsequently available sets of decisions. The state space of a non-deterministic finite automaton is evolving due to the computation itself hence it cannot be efficiently contracted using a deterministic finite automaton that will arrive at a solution in super-polynomial time. The solution of the $\mathcal{NP}$ problem itself is verifiable in polynomial time ($\mathcal{P}$) because the corresponding state is stationary. Likewise the class $\mathcal{P}$ set of states does not depend on computational history hence it can be efficiently contracted to the accepting state by a deterministic sequence of dissipative transformations. Thus it is concluded that the class $\mathcal{P}$ set of states is inherently smaller than the set of class $\mathcal{NP}$. Since the computational time to contract a given set is proportional to dissipation, the computational complexity class $\mathcal{P}$ is a subset of $\mathcal{NP}$.

## 1. Introduction

The distinction between computational complexity classes referred to as $\mathcal{P}$ and $\mathcal{NP}$ has remained ambiguous (1,2). On one hand, decision problems in class $\mathcal{P}$ can be solved efficiently by a deterministic algorithm within a number of steps bound by a polynomial function of the length of the input. An example of a $\mathcal{P}$ problem is that of the shortest path: what is the least-cost one-way path through a given network of cities to the destination? On the other hand, to solve problems in class $\mathcal{NP}$ efficiently seems to require some non-deterministic parallel machine, yet solutions can be verified as correct in a deterministic manner. An example of a $\mathcal{NP}$-complete problem is that of the traveling salesman: what is the least-cost round-trip path via a given network of cities, visiting each exactly once?

The ambiguity between classes $\mathcal{P}$ and $\mathcal{NP}$ prevails because it appears, although it has not been proven, that the traveling salesman problem (3) and numerous other $\mathcal{NP}$ problems in mathematics, physics, biology, economics, optimization, artificial intelligence, *etc.*, (4) cannot be solved in polynomial time by deterministic finite automata unlike the shortest path problem and other $\mathcal{P}$ problems. Yet, the initial instances of the traveling salesman and the shortest path problem seem to differ at most polynomially from one another. Therefore, could it be that there are, after all, for the $\mathcal{NP}$ problems as efficient algorithms as there are for the $\mathcal{P}$ problems but these simply were not found yet?

In this study insight to the $\mathcal{P}$ *versus* $\mathcal{NP}$ question is acquired from the 2$^{nd}$ law of thermodynamics (5,6,7). The natural law was recently written as an equation of motion and associated with the principle of least action and Newton's second law (8,9,10). The old ubiquitous imperative, known also as the principle of increasing entropy, describes a system in evolution toward more probable states. Here, it is of particular interest that evolution is in general a non-deterministic process as is class $\mathcal{NP}$ computation. Furthermore, the end point of evolution, *i.e.*, the stable stationary state itself can be efficiently validated as the free energy minimum in a similar manner as the solution to a $\mathcal{NP}$ computation can be verified as accepting.

The recent formulation of the 2$^{nd}$ law as an equation of motion based on statistical mechanics of open systems has rationalized diverse evolutionary courses that result in skewed distributions whose cumulative curves are open-form integrals (11,12,13,14,15,16,17,18,19). Some of these natural processes (20), *e.g.*, protein folding that directs down along intractable trajectories to diminish free energy (21), have been recognized as the hardest problems in class $\mathcal{NP}$ (22). Although, many other $\mathcal{NP}$-complete problems do not apparently concern physical reality, the concept of $\mathcal{NP}$-completeness (23) encourages one to consider computation as an energy transduction process that follows the 2$^{nd}$ law. The physical portrayal of computation allows one to use the fundamental theorems concerning conserved currents (24) and gradient systems (20,25) in the classification of computational complexity. Specifically, it is found that the circuit currents remain tractable during the class $\mathcal{P}$ problem computation because the accessible states do not dependent on the processing steps themselves. Thus the class $\mathcal{P}$ state set can be efficiently contracted using a deterministic finite automaton to the accepting set along the dissipative path without additional degrees of freedom. In contrast, the circuit currents are intractable during the class $\mathcal{NP}$ problem computation because each step of the problem-solving process depends on the computational history and affects future decisions. Thus the contraction of states along alternative but interdependent paths to the accepting set is a non-deterministic process.

The adopted physical perspective on computation is consistent with the standpoint that no information exists without its physical representation (26,27) and that information processing itself is governed by the $2^{nd}$ law (28). The connection between computational complexity and the natural law also yields insight to the abundance of natural problems in class $\mathcal{NP}$ (4). In the following, the description of computation as an evolutionary process is first outlined and then developed to mathematical forms to make the distinction between the classes $\mathcal{P}$ and $\mathcal{NP}$.

## 2. Computation as a physical process

According to the $2^{nd}$ law of thermodynamics a computational circuit, just as any other physical system, evolves by diminishing energy density differences within the system and relative to its surroundings. The energy dispersal process (29) is generally referred to as evolution. Flows of energy naturally select (30) the steepest descents in the free energy landscape to abolish the density differences as soon as possible (10). A clocked circuit as a physical realization of a finite automaton is an energy transduction network. In accordance with the network notion, the $\mathcal{P}$ vs. $\mathcal{NP}$ question can be phrased in terms of graphs (31) that are networks of Boolean components and shift register nodes.

Computation is, according to the principle of increasing entropy, a probable process. It will begin when an energy density difference, representing an input, appears at the interface between the computational system and its surroundings. Thus, the input places the automaton at the initial state of evolution. A specific input string of alphabetic symbols is represented to the circuit by a particular physical influx, *e.g.*, as a train of voltages. No instance is without physical realization.

The algorithmic execution is an irreversible thermalization process where the energy absorbed at the input interface will begin to disperse within the circuit. Eventually, after a series of dissipative transformations from a state to another, more probable one, the computational system arrives at a thermodynamic steady state, the final acceptance, by emitting an output, *e.g.*, as the tape stops writing a solution. No solution can be produced without physical representation.

Physically speaking, the most effective problem solving is about finding the path of least action, which is equivalent to the maximal energy transduction from the initial instance down along the steepest gradients in free energy to the final acceptance. However, the path for the optimal conductance, *i.e.*, for the most rapid reduction of free energy, is tricky to find in a circuit with three or more degrees of freedom because flows (currents) and forces (voltages) are inseparable. In contrast, when the process has no additional degrees of freedom in dissipation, the minimal resistance path corresponding to the solution can be found in a deterministic manner.

In the general case the path is intractable because the state space keeps changing due to the search itself. The decision to move from the present state to another depends on the past decisions and will also affect accessible states in the future. For example, when the traveling salesman decides for the next destination, the decision will depend on the past path, except at the very end, when there are no choices but to return home. The path is directed because revisits are not allowed (or eventually restricted by costs). This class, referred to as $\mathcal{NP}$, contains intractable problems that describe irreversible (directional) processes (Fig. 1) with additional ($n \geq 3$) degrees of freedom.

In the special case the path is tractable as decisions are independent of computational history. For example, when searching for the shortest path through a network, the entire invariant state space is, at least in principle, visible from the initial instance, *i.e.*, the problem is deterministic. A decision at any node is independent of traversed paths. This class, referred to as $\mathcal{P}$, contains tractable problems that describe irreversible processes without additional degrees of freedom. Moreover, when the search among alternatives is not associated with any costs, the process is reversible (non-directional), *i.e.*, indifferent to the total conductance from the input to output node that is to be maximized.

Finally, it is of interest to note that a particular physical system may have no mechanisms to proceed from a state to any other by transforming absorbed quanta to any emission. Since dispersion relations of physical systems are revealed first when interacting with them (32,33), it is impossible to know for a given circuit and finite influx, *a priori*, without interacting whether the system will arrive at the free energy minimum state finishing with emission or remain at an excited state without output forever. This is the physical rationale of the halting problem (34). It is impossible to decide for a given program and finite input, *a priori*, without processing whether the execution will arrive at the accepting state finishing with output or remain at a running state without output forever. These processes that acquire but do not yield, relate to problems that cannot be decided. They are beyond class $\mathcal{NP}$ (35) and will not be examined further. Here the focus is on the principal difference between the truly tractable and inherently intractable problems.

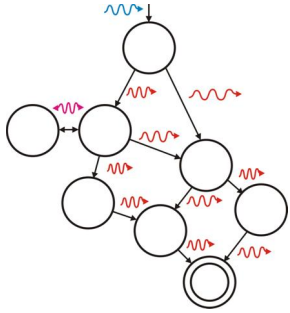

Figure 1. During computation an influx of energy disperses from the input interface (top) through the network that evolves, by dissipative transitions that acquire (blue) and yield (red) energy, toward the stationary state (bottom). Reversible transitions, *i.e.*, conserved currents (purple), do not bring about changes of state and advance the computation. Driving forces (free energy between the nodes) and flows (between the nodes) are inseparable when there are additional degrees of freedom ($n \geq 3$), *i.e.*, alternative but interdependent paths for the dissipative processes to proceed along. Then the flows are intractable and the corresponding algorithmic execution is non-deterministic.

## 3. Self-similar circuits

The physical portrayal of problem processing according to the principle of increasing entropy is based on the hierarchical and holistic formalism (36). It recognizes that circuits are self-similar in their functional organization (Fig. 2) (16,37,38). A circuit is composed of circuits, or equivalently, there are networks within nodes of networks.

Each node of a transduction network is a physical entity associated with energy $G_k$. A set of identical nodes $N_k > 0$ representing, for example, a memory register, is associated, following Gibbs (39), with a density-in-energy defined by $\phi_k = N_k \exp(G_k/k_B T)$ relative to the average energy density $k_B T$. The self-similar formalism assigns to a set of indistinguishable nodes in numbers $N_k$ a probability measure $P_k$ (8,40)

$$P_k = \left[ \prod_n N_n \exp\left( \frac{-\Delta G_{kn} + \Delta Q_{kn}}{k_B T} \right) \Big/ g_{kn}! \right]^{N_k} \Big/ N_k! \quad (3.1)$$

in a recursive manner, so that each node $k$ in numbers $N_k$ is a product of embedded $n$-nodes, each distinct type available in numbers $N_n$. The combinatorial configurations of identical $n$-nodes in the $k$-node are numbered by $g_{kn}$. Likewise, the identical $k$-nodes in numbers $N_k$ are indistinguishable from each other in the network. The internal difference $\Delta G_{kn} = G_k - G_n$ and the external flux $\Delta Q_{kn}$ denote the quanta of (interaction) energy.

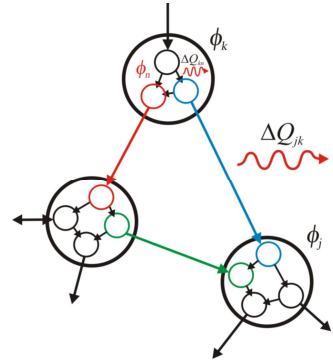

Figure 2. Network nodes are networks themselves according to the self-similar formulation of energy transduction. Any two densities $\phi_j$ and $\phi_k$ at the nodes $j$ and $k$ are distinguished from each other by a dissipative $jk$-transformation $\Delta Q_{jk} \neq 0$.

The computational system is processing from a state to another more probable one when energy is flowing down along gradients through the network from a node to another with concurrent dissipation to the surroundings. For example, a $j$-node can be driven from its present state, defined by the potential $\mu_j = k_B T \ln \phi_j$ (29), to another state by an energy flow from a preceding $k$-node at a higher potential $\mu_k$ and by an energy efflux $\Delta Q_{jk}$ to the surroundings (Fig. 2). Subsequently the $j$-node may transform anew from its current high-energy state to a stationary state by yielding an efflux to a connected $i$-node at a lower potential coupled with emission to the surroundings. Any two states are distinguished from each other as different only when the transformation from one to the other is dissipative $\Delta Q_{jk} \neq 0$ (8,9,10). When thermalization has abolished all density differences, the irreversible process has arrived at a dynamic steady state where reversible, to-and-fro flows of energy (currents) are conserved and, on the average, the densities remain invariant.

It is convenient to measure the state space of computation by associating each *j*-system with logarithmic probability

$$\ln P_j \approx N_j\left(1 - \sum_k \frac{\Delta\mu_{jk} - \Delta Q_{jk}}{k_B T}\right) = N_j\left(1 - \sum_k \frac{\Delta V_{jk}}{k_B T}\right) \quad (3.2)$$

in analogy to Eq. 3.1 where $\Delta\mu_{jk}/k_B T = \ln\phi_j - \Sigma\ln(\phi_k/g_{jk}!)$ is the potential difference between the *j*-node and all other connected *k*-nodes in degenerate (equal-energy) numbers $g_{jk}$. Stirling's approximation implies that $\ln P_j$ is a sufficient statistic (41) for $k_B T$ so that the system may accept (or discard) quanta without marked changes in its total energy content, *i.e.*, the free energy $\Delta V_{jk} = \Delta\mu_{jk} - \Delta Q_{jk} \ll k_B T$. Otherwise, a high influx $\Delta V_{jk} \gtrsim k_B T$, such as a voltage spike from the preceding *k*-node or heat from the surroundings, might "damage" the *j*-system, *e.g.*, "burn" a memory register, by forcing the embedded *n*-nodes into evolution (Fig. 2). Such a non-statistic phenomenon may manifest itself even as chaotic motion but this is no obstacle for the adopted formalism. Then the same self-similar equations are used at a lower level of hierarchy to describe processes involving sufficiently statistic systems.

According to the scale-independent formalism the network is a system in the same way as its constituent nodes are systems themselves. Any two networks, just as any two nodes, are distinguishable from each other when there is some influx sequence so that exactly one of the two systems is transforming. In computational terms, any two states of a finite automaton are distinguishable when there is some input string so that exactly one of the two transition functions is accepting (2). Those nodes that are distinguishable from each other by mutual density differences are non-equivalent. These distinct fractions of a circuit are represented by disjoint sets and indexed separately in the total additive measure of the entire circuit defined as

$$\ln P = \sum_{j=1} \ln P_j = \sum_{j=1} N_j\left(1 - \sum_{k \neq j} \frac{\Delta V_{jk}}{k_B T}\right). \quad (3.3)$$

The affine union of disjoint sets is depicted as a graph that is merged from subgraphs by connections.

The logarithmic measure $\ln P$ (Eq. 3.3) implies a complicated energy transduction network by its indexing of numerous nodes as well as differences between them and in respect to the surroundings. In a sufficiently statistical system the changes in occupancies balance as $\Delta N_j = -\Sigma\Delta N_k$. The influx to the *j*-node results from the effluxes from the *k*-nodes (or *vice versa*). The flows along the *jk*-edges are proportional to the free energy by invariant conductance $\sigma_{jk} > 0$ defined as (8)

$$\Delta N_j = -\sum_k \sigma_{jk} \frac{\Delta V_{jk}}{k_B T}. \quad (3.4)$$

The form ensures continuity so that when a particular *jk*-flow is increasing the occupancy $\Delta N_j > 0$ of the *j*-node, the very same flow is decreasing the occupancies $\Sigma\Delta N_k < 0$ at the *k*-nodes (or *vice versa*). Importantly, owing to the other affine connections, the *jk*-transformation will affect occupancies of other nodes that in turn affect $\Delta V_{jk}$. Consequently when there are, among interdependent nodes ($n \geq 3$), alternative paths ($k \geq 2$) of conduction, the problem of finding the optimal path becomes intractable (8,10). As long as $\Delta V_{jk} \neq 0$ the gradient system with $n \geq 3$ degrees of freedom does not enclose integrable (tractable) orbits (25).

Conversely in the special case, when the reduction of a difference does not affect other differences, *i.e.*, there are no additional degrees of freedom, the changes in occupancies remain tractable. The conservation of energy requires that when there are only two degrees of freedom, the flow from one node will inevitably arrive exclusively at the other node. Therefore it is not necessary to explore all these integrable paths to their very ends as the outcome can be predicted and the particular path in question can be found efficiently. Moreover, when there are no differences $\Delta V_{jk} = 0$, there are no net variations, *i.e.*, no net flows either. These conserved, reversible flows are statistically predictable even in a complicated but stationary ($\Delta\ln P = 0$) network with degrees of freedom. When the currents are conserved, the network is idle, *i.e.*, not transforming. In accordance with Noether's theorem also the Poincaré-Bendixson theorem holds for the stationary system (20,25).

The overall transduction processes, both intractable and tractable, direct toward more probable states, *i.e.*, $\Delta\ln P > 0$. However when a natural process with three or more degrees of freedom is examined in a deterministic manner, it is necessary to explore all conceivable transformation paths to their ends. The paths cannot be integrated in closed forms (predicted) because each decision will affect the choice of future states. The set of conceivable states that is generated by decisions at consequent branching points of computation can be enormous.

The physical portrayal of computational complexity reveals that it is the non-invariant, evolving state space of class $\mathcal{NP}$ that prevents from completing the contraction by dissipative transformations in polynomial-time. Since the dissipated flow of energy during the computation relates directly to the irreversible flow of time (10), the class $\mathcal{NP}$ completion time is inherently longer than that of class $\mathcal{P}$. Thus it is concluded that $\mathcal{P}$ is a subset of $\mathcal{NP}$.

### 4. Computation as a probable process

During the probable physical process of computation the additive logarithmic probability measure $\ln P$ is increasing when the dissipative transformations are leveling the differences $\Delta V_{jk} \neq 0$ ($\Delta V_{jj} = 0$). When the definitions in Eq. 3.4 and $\Delta \mu_{jk}(\Delta N_j)/k_B T = \Delta N_j/N_j$ are used, the change $\Delta \ln P$

$$L = \Delta \ln P = -\sum_{j=1} \Delta N_j \sum_k \frac{\Delta V_{jk}}{k_B T} = \sum_{j,k} \sigma_{jk} \left(\frac{\Delta V_{jk}}{k_B T}\right)^2 \geq 0 \quad (4.1)$$

is found to be non-negative since the squares $(\Delta V_{jk})^2$ and $(\Delta N_j)^2$ are necessarily non-negative and the absolute temperature $T > 0$, $\sigma_{jk} \geq 0$ and $k_B > 0$.

The definition of entropy $S = k_B \ln P$ yields from Eq. 4.1 the principle of increasing entropy $\Delta S = -\Sigma_j \Delta N_j \Sigma_k \Delta V_{jk}/T \geq 0$. Equation 4.1 says that entropy is increasing when free energy is decreasing, in agreement with the thermodynamic maxim (29) and Gouy-Stodola theorem (42,43) and the mathematical foundations of thermodynamics (44,45,46). In other words, when the process generator $L > 0$, there is free energy for the computation to commence from the initial state toward the accepting state where the output thermalizes the circuit and $L = 0$. Admittedly, dissipation is often small, however, not negligible but necessary for any computation to yield an output (26,27,28).

During the computational process the state space accessible by $L > 0$ is contracting toward the free energy minimum state where $L = 0$ and no further changes of state are possible. Consistently, when $\ln P$ is increasing due to the changing occupancies $\Delta N_j$, the change in the process generator (20)

$$\Delta L = 2\sum_{j=1} \frac{\Delta N_j}{N_j} \sum_k \sigma_{jk} \frac{\Delta V_{jk}}{k_B T} = -2\sum_{j=1} \frac{(\Delta N_j)^2}{N_j} \leq 0 \quad (4.2)$$

is found to decrease almost everywhere using the definition in Eq. 3.4 because the squares $(\Delta N_j)^2$ and $(\Delta V_{jk})^2$ are necessarily non-negative and $N_j > 0$ for any spatially confined energy density (10). The equations 4.1 and 4.2 show that during the computation the state space is contracting toward the stationary state where $L = 0$.

The free energy minimum partition $\ln P_{max} = \Sigma N_j^{ss}$ corresponding to the solution is stable in its surroundings because any variation $\delta N_j$ below (above) the steady-state occupancy $N_j^{ss}$ will reintroduce $\Delta V_{jk} < 0$ ($> 0$) that will drive the system back to the stationary state in its surroundings by invoking a returning flow $\Delta N_j > 0$ ($< 0$). Explicitly, the maximum entropy system is Lyapunov stable (20,25) according to the definitions $\delta \ln P = L(\delta N_j) < 0$ and $\delta L(\delta N_j) > 0$ available from Eqs. 4.1 and 4.2. The dynamic steady state is maintained by frequent to-and-fro flows (interactions) between the system's constituents and the surroundings. Moreover, non-dissipative processes do not amount to any change in $P$.

In general, the trajectories of natural processes cannot be solved analytically because the flows $\Delta N_j$ and $\Delta V_{jk}$ are inseparable in $L$ (Eq. 4.1) at any $j$-node where cardinality of $\{j,k\} \geq 3$. The inherently intractable processes can be simulated by updating $T$, $\Delta V_{jk}$ and $N_j$ after each change of state. The occupancies $N_j$ keep changing due to the changing driving forces $\Delta V_{jk}$ that, in turn, are affected by the changes $\Delta N_j$. The non-Hamiltonian system is without invariants of motion and Liouville's theorem is not satisfied because the open dissipative system is subject to an influx (efflux) from (to) its surroundings. The non-conserved, gradient system is without norm and the evolving (*cf.* Bayesian) distribution of probabilities $P_j$ cannot be normalized. The dissipative equation of motion $\Delta P/\Delta t = LP$ for the class of irreversible processes cannot be integrated in a closed form or transformed to a time-independent frame (10) to obtain a solution efficiently.

According to the maximum entropy production principle (47,48,49,50,51,52,53,54,55,56,57,58,59) the energy differences are reduced most effectively when entropy increases most rapidly, *i.e.*, most voluminous currents direct along the steepest paths. However, when choosing at every instance a particular descent that appears as the steepest, there is no guarantee that the most optimal path will be found because the transformations themselves will affect the future states between the initial instance and the final acceptance. To be sure about the optimal trajectory it takes time (dissipation) because the deterministic algorithmic execution of the class $\mathcal{NP}$ problem will have to address by conceivable transformations the entire power set of states, one member for each distinct path of energy dispersal.

In the special case when the currents are separable from the driving forces, the energy transduction network remains invariant. The Hamiltonian system has invariants of motion and Liouville's theorem is satisfied. The deterministic computation as a tractable energy transduction process will solve the problem in question because the dissipative steps are without additional degrees of freedom. The conceivable courses can be integrated (predicted). Hence the solution can be obtained efficiently, *e.g.*, by an algorithm that follows the steepest descent and does not waste time in wandering along paths that can be predicted to be futile.

## 5. Manifold in motion

Further insight to the distinction between computations in the classes $\mathcal{P}$ and $\mathcal{NP}$ is obtained when the computation as a physical process is described in terms of an evolving energy landscape (60,61,62). To this end the discrete differences $\Delta$ that denote properly transforming forces and quantized flows, are replaced by differentials $\partial$ of continuous variables. A spatial gradient $\partial U_{jk}/\partial x_j$ is a convenient way to relate a density labeled by $j$ at a continuum coordinate $x_j$ with another one labeled by $k$ but displaced by dissipation $\partial Q_{jk}/\partial t$ at $x_k$ (9,10). When the $j$-system at $x_j$ evolves down along the scalar potential gradient $\partial U_{jk}/\partial x_j$ in the field $\partial Q_{jk}/\partial x_j$, the conservation of energy requires that the transforming current $v_j = dx_j/dt = -\Sigma dx_k/dt$. The radiated dissipation $\partial Q_{jk}/\partial t$ is an efflux of photons at the speed of light $c$ to the surrounding medium (or *vice versa*).

The continuum equation of motion corresponding to Eq. 4.1 is obtained from Eq. 3.3 by differentiating and using the chain rule $(dP_j/dx_j)(dx_j/dt)$ (10)

$$L = -\sum_{j,k} D_j \frac{V_{jk}}{k_B T} \quad (5.1)$$

where directional derivates $D_j = (dx_j/dt)(\partial/\partial x_j)$ span the affine manifold (63) of energy densities (Fig. 3). The total potential $V_{jk} = U_{jk} - iQ_{jk}$ is decomposed to the orthogonal scalar $U_{jk}$ and dissipative $Q_{jk}$ parts (64). All distinguishable densities and flows are indexed by $j \neq k$. The evolving energy landscape is concisely given by the total change in kinetic energy $\partial(2K)/\partial t = k_B T L = T \partial S/\partial t$ (9,10)

$$\sum_{j,k} v_j \frac{\partial}{\partial t} m_{jk} v_k = \sum_{j,k} v_j m_{jk} \frac{\partial v_k}{\partial t} + \sum_{j,k} v_j \frac{\partial m_{jk}}{\partial t} v_k \quad (5.2)$$
$$\Leftrightarrow \frac{\partial}{\partial t} 2K = -\sum_{j,k} v_j \frac{\partial U_{jk}}{\partial x_j} + \sum_{j,k} \frac{\partial Q_{jk}}{\partial t}$$

where three or more degrees of freedom ($n \geq 3$) are denoted by indexing $j \neq k \pm 1$. Conversely, the lack of additional degrees of freedom ($n < 3$) is indicated by indexing $j = k \pm 1$.

The equation for the flows of energy can also be obtained from Newton's 2$^{nd}$ law (65) for the change in momentum $p_{jk} = m_{jk} v_k$

$$\frac{\partial}{\partial t} \sum_{j,k} p_{jk} = \sum_{j,k} m_{jk} a_k + \sum_{j,k} \frac{\partial m_{jk}}{\partial t} v_k \quad (5.3)$$
$$= -\sum_{j,k} \frac{\partial V_{jk}}{\partial x_j} = -\sum_{j,k} \frac{\partial U_{jk}}{\partial x_j} + \sum_{j,k} \frac{\partial Q_{jk}}{v_j \partial t}.$$

by multiplying with velocities. The gradient $\partial V_{jk}/\partial x_j$ is again decomposed to the spatial and temporal parts. The sign convention is the same as above, *i.e.*, when $\partial U_{jk}/\partial x_j < 0$, then $v_j > 0$. Since momenta are at all times tangential to the manifold, Newton's 2$^{nd}$ law (Eq. 5.3) requires that the corresponding flow at any moment

$$v_j = -\sum_k \frac{\sigma_{jk}}{k_B T} \frac{\partial V_{jk}}{\partial x_j} \quad (5.4)$$

is proportional to the driving force in accordance with the continuity $v_j = -\Sigma v_k$ across the $jk$-edges between nodes of the network (Eq. 3.4) (8). The linear relationship in Eq. 5.4 that is reminiscent of Onsager reciprocal relations (44), is consistent with the previous notion that the densities-in-energy (the nodes) are sufficiently statistic. Otherwise, a high current between $x_k$ and $x_j$ would force the underlying conducting system ($jk$-edge), parameterized by the coefficient $\sigma_{jk}$, to evolution. In such a case the channel (conductance) characteristics would depend on transmitted bits (28).

A particular flow $v_j$ funnels by dissipative transformations down along the steepest descent $-\partial V_{jk}/\partial x_j$, *i.e.*, along the shortest path $s_{jk} = \int d(\sqrt{v_j m_{jk} v_k})$ known as the geodesic (44). At any given moment the positive definite resistance $r_{jk} = k_B T \sigma_{jk}^{-1} > 0$ in Eq. 5.4 identifies to the mass $m_{jk} > 0$ that defines the geometry of the free energy landscape (66) (*cf.* Lorentzian manifold). Formally $s_{jk}$ can be denoted as an integral, however in the general case of the evolving non-Euclidean landscape it cannot be integrated in a closed form (25). The curved landscape is shrinking (or growing) because the surroundings are draining it by a net efflux (or supplying it with a net influx) of radiation $\partial Q_{jk}/\partial t \neq 0$ and/or a material flow $\partial U_{jk}/\partial t \neq 0$. When the forces and flows are inseparable in $L$, the non-invariant landscape is, at any given locus and moment, a result of its evolutionary history. The rate of net emission (or net absorption) declines as the system steps, quantum by quantum, toward the free energy minimum, which is the stationary state in the respective surroundings. Only in the special case, when the forces and flows are separable, can the trajectories be integrated in a closed form.

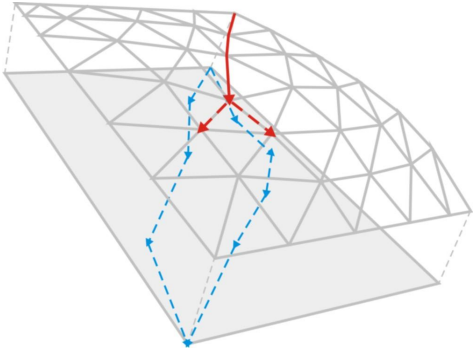

Figure 3. The curved energy landscape, covered by triangles, represents the state set of intractable computation. The non-Euclidian manifold is evolving by the contraction process itself toward the optimal path of maximal conduction (red arrows) corresponding to the solution. During the contraction the path from the initial instance (top) toward the final acceptance (bottom) is shortening but remains non-integrable (unpredictable) due to the dissipation with additional degrees of freedom (exemplified at a branching point). In contrast the paths (blue arrows) on the invariant Euclidean plane (grey) do not mold the landscape and thus they do not have to be followed but can be integrated (predicted).

Finally, when all density differences have vanished, the manifold has flattened to the stationary state ($dS/dt = 0$). The state space has contracted to a single stationary state where $L = 0$. In agreement with Noether's theorem the currents are conserved and tractable throughout the invariant manifold. Also in accordance with Poincaré's recurrence theorem the steady-state reversible dynamics are exclusively on bound and (statistically) predictable orbits. Moreover the conserved currents, i.e., $\partial m_{jk}/\partial t = 0$, bring about no net changes in the total energy content of the system. Hence Eq. 5.3 reduces to

$$\sum_{j,k} v_j \frac{\partial}{\partial t} m_{jk} v_k = \sum_{j,k} v_j m_{jk} \frac{\partial v_k}{\partial t} \Leftrightarrow \frac{\partial}{\partial t} 2K = -\sum_{j,k} v_j \frac{\partial U_{jk}}{\partial x_j} \quad (5.5)$$

which implies in accordance with the virial theorem that the components of kinetic energy $2K$ match the components of potential $U$ everywhere.

According to the geometric description of computational processes, the flattening (evolving) non-Euclidean landscape represents the state space of the class $\mathcal{NP}$ computation whereas the flat Euclidean manifold represents the state space of the class $\mathcal{P}$ computation. The geodesics that span the class $\mathcal{NP}$ landscape are arcs whereas those that span the class $\mathcal{P}$ manifold are straight lines. According to Eq. 5.2 the class $\mathcal{NP}$ state space is, due to its three or more degrees of freedom ($n \geq 3$), larger in dissipation by the terms $\Sigma v_j dm_{jk} v_k > 0$ indexed with $j \neq k \pm 1$, than the class $\mathcal{P}$ state space without additional degrees of freedom ($n < 3$) for dissipation given by the term $\Sigma v_j dm_{jk} v_k > 0$ indexed with $j = k \pm 1$. In other words, class $\mathcal{NP}$ is larger than $\mathcal{P}$ because the curved manifold cannot be embedded in the plane. The measure $\ln P_{\mathcal{NP}}$ of the non-Euclidean landscape is simply larger by the degrees of freedom ($n \geq 3$) in dissipation than the measure $\ln P_{\mathcal{P}}$ of Euclidean manifold without additional degrees of freedom.

The argument for the failure to map the larger $\mathcal{NP}$ manifold one-to-one onto the smaller $\mathcal{P}$ manifold is familiar from the pigeonhole principle $PHP^{\mathcal{NP}}_{\mathcal{P}}$ applied for manifolds $\ln P_{\mathcal{NP}} > \ln P_{\mathcal{P}}$. The quanta that are dissipated during evolution from diverse density loci of the curved, evolving $\mathcal{NP}$ landscape are not mapped anywhere on the flat, invariant $\mathcal{P}$ landscape. Thus it is concluded that $\mathcal{P}$ is a subset of $\mathcal{NP}$.

## 6. Intractability in the degrees of freedom

The transduction path between two nodes can be represented by only one edge, hence there are $k = n - 1$ interdependent currents (Eq. 3.4) between $n$ densities (20). The degrees of freedom are less than $n$ by 1 because it takes at least two densities to have a difference. In the general case $n \geq 3$, there are alternative paths for the currents from the initial state via alternative states toward the accepting state. The intractable evolutionary courses are familiar from the $n$-body ($n \geq 3$) problems (67,68). Accordingly, the satisfiability problem of a Boolean expression ($n$-SAT) belongs to class $\mathcal{NP}$ when there are three or more literals ($n \geq 3$) per clause (23). In the special case $n = 2$, the energy dispersal process is deterministic as there are no alternative dissipative paths for the current. When only one path is conducting, the problem for the maximal conduction is 1-separable and tractable. The two-body problem does not present a challenge. Accordingly, 2-SAT is deterministic and 1-SAT is trivial, essentially only a statement.

For example, the problem of maximizing the shortest path by two or more interdicts ($k \geq 2$) is intractable. When the first interdict is placed, flows are redirected and, in turn, affect the decision to place the second interdict. Similarly the search history of the traveling salesman for the optimal round-trip path is intractable. A decision to visit a particular city will narrow irreversibly the available state space by excluding that city from the subsequent choices. Thus, at any particular node one cannot consider decisions as if not knowing the specific search history that led to that node. When each decision will open a new set for future decisions, the computational space state of class $\mathcal{NP}$ is a tedious power set of deterministic decisions. On the other hand when optimizing the shortest path, a choice for a particular path does not affect, in any way, the future explorations of other paths. At any particular node one may consider decisions irrespective of the search history. In the deterministic case it is not necessary to explore all conceivable choices because the trajectories are tractable (predictable). Likewise, the problem of maximizing the shortest path by a single interdict $k = 1$ can be solved efficiently. Any particular decision to place the interdict does not affect future decisions because there are no more interdicts to be placed. When the state space is not affected by the problem-solving process itself, at most, a polynomial array of invariant circuits, *i.e.*, deterministic finite automata, will compute class $\mathcal{P}$ problems.

The $\mathcal{P}$ *vs.* $\mathcal{NP}$ question is not only a fundamental but also a practical problem for which no computational machinery exists without physical representation. A particular input instance is imposed on the computational circuit by the surroundings and a particular output is accepted as a solution by the surroundings. The communication between the automaton and its surroundings relates to information processing that was understood already early on to be equivalent to the (impedance) matching of circuits for optimal energy transmission (69). When the matching of a circuit will affect the matching of two or more connected circuits, the total matching of the interdependent circuits for the optimal overall transduction is intractable. Although in practice the iterative process may be converging rapidly in a non-deterministic manner, the conceivable set of circuit states is a power set of the tuning operations. Conversely, when the matching does not involve degrees of freedom, the tuning for optimal transduction is tractable.

In summary, the class $\mathcal{NP}$ problem solving process is inherently non-deterministic because the contraction process will itself affect the set of future states accessible from a particular instance. The course toward acceptance cannot be accelerated by prediction but the state space must be explored. On the other hand when dissipative steps between the input and output operations have no additional degrees of freedom, the search for the class $\mathcal{P}$ problem solution will itself not affect the accessible set of states at any instance. The invariant state set can be contracted efficiently by predicting rather than exploring all conceivable paths. Therefore, the completion time of the class $\mathcal{P}$ deterministic computation is shorter than that of $\mathcal{NP}$. Thus it is concluded that $\mathcal{P}$ is a subset of $\mathcal{NP}$.

## 7. State spaces of automata

The computational complexity classification to $\mathcal{P}$ and $\mathcal{NP}$ by the differing degrees of freedom in dissipation relates to the algorithmic execution times, which are proportional to circuit sizes. A Boolean circuit that simulates a Turing machine is commonly represented as a (directed, acyclic) graph structure of a tree with the assignments of gates (functions) to its vertices (nodes) (Fig. 2).

The class $\mathcal{NP}$ problems are represented by circuits where forces (voltages) are inseparable from currents. Since there are no invariants of motions, the *ceteris paribus* assumption does not hold when solving the class $\mathcal{NP}$ problems (70). Consistently, no deterministic algorithms are available for the class of non-conserved flow problems but, *e.g.*, brute-force optimization, simulated annealing and dynamic programming are employed (71).

The class $\mathcal{NP}$ problems can be considered to be computed by a non-deterministic finite automaton (NFA). It is a finite state machine where for each pair of state and input symbol there may be several possible states to be accessed by a subsequent transition. The NFA 5-tuple ($\Phi$, $\Delta$, $\Lambda$, $\phi_1$, $\phi_{ss}$) consists of a finite set of states $\Phi$, a finite set of input symbols $\Delta$, a transition function $\Lambda: \Phi \times \Delta \rightarrow P(\Phi)$, where $P(\Phi)$ denotes the power set of $\Phi$, an initial state $\phi_1 \in \Phi$, a set of accepting (stationary) states $\phi_{ss} \subseteq \Phi$.

A circuit for the non-deterministic computation can also be constructed from an array of deterministic finite automata (DFA). Each DFA is a finite state machine where for each pair of state and input symbol there is one and only one transition to the next state. The DFA 5-tuple ($\Phi$, $\Delta$, $\Lambda$, $\phi_1$, $\phi_{ss}$) consists of a finite set of states ($\Phi$), a finite alphabet $\Delta$, a transition function $\Lambda: \Phi \times \Delta \to \Phi$, an initial state ($\phi_1 \in \Phi$), a set of accepting states ($\phi_{ss} \subseteq \Phi$).

In the general case when the forces are inseparable from flows, the execution time by the DFA array grows super-polynomial as function of the input length $n$, *e.g.*, as $O(N^n)$. For example, when maximizing the shortest path by interdicts ($k \geq 2$), any two alternative choices will give rise to two circuits that differ from each other as much as the currents of the two DFAs differ from each other. These two sets are non-equivalent due to the difference in dissipation, and one cannot be reduced to the other. Accordingly, the circuit for the NFA is adequately constructed from the entire power set of distinct DFAs to cover the entire conceivable set of states of the non-deterministic computation (Fig. 4). The union of DFAs is non-reducible, *i.e.*, each DFA is distinguished from all other DFAs by its distinct transition function.

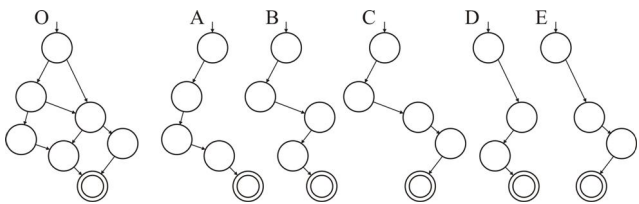

Figure 4. A circuit (O) containing nodes with degrees of freedom ($n \geq 3$) represents an NFA. The computation steps from a state to another when currents are driven from the input instance (top) down along alternative but interdependent paths toward the output acceptance (bottom). Since the currents affect each other by affecting the driving forces, the circuit corresponds to the NFA having a power set of states. It can be decomposed to the distinct circuits (A – E), one member for each conceivable current without additional degrees of freedom, that are representing an array of DFAs each having at most a polynomial set of states.

The class $\mathcal{P}$ problems are represented by circuits where forces are separable from currents. When the proposed questions do not depend on previous decisions (answers), the problem can be computed efficiently by DFA. Consistently in the class $\mathcal{P}$ of flow conservation problems many methods deliver the solution corresponding to the maximum flow in polynomial time. For example, during the search for the maximally conducting path through the network, currents disperse from the input node $k$ to diverse alternative nodes $l$ but only the flow along the steepest descent arrives at the output node $j$ and establishes the only and most voluminous flow. The other paths of energy dispersal terminate at dead ends and do not contribute or affect the maximum flow at all. Importantly, on an invariant landscape these inferior paths do not have to be followed to their very ends as is exemplified by Dijkstra's algorithm (72). The search terminates at the accepting state whereas other paths end up at nil states. These particular sequences of states "died". The shortest path problem can be presented by a single DFA because the non-accepting dead states that keep going to themselves, belong to $\varnothing$, the empty set of states. However, as has been accurately pointed out (2), technically this automaton is a non-deterministic finite automaton, which reflects understanding that the single flow without additional degrees of freedom ($n = 2$) is the special deterministic subclass of the generally ($n \geq 3$) non-deterministic class. Likewise, the special case of maximizing the shortest path by a single interdict ($k = 1$) is deterministic in contrast to the general case of two or more interdicts ($k \geq 2$). The special 1-separable problem can be represented by a linear set of distinct circuits in contrast to the general inseparable problem that requires a power set of distinct circuits. Accordingly, the automaton for the special cases of deterministic problems is adequately constructed at most from a polynomial set of distinct DFAs and the corresponding computation is completed in polynomial time.

Since the class $\mathcal{NP}$ varying state space is larger, due to its additional degrees of freedom, than the class $\mathcal{P}$ invariant state space, it is concluded that $\mathcal{P}$ is a subset of $\mathcal{NP}$.

## 8. The measures of states

To measure the difference between the classes $\mathcal{P}$ and $\mathcal{NP}$, the thermodynamic formalism of computation will be transcribed to the mathematical notation (45). Consistently with the reasoning presented in sections 2 – 7, the computational complexity class $\mathcal{P}$ will be distinguished from $\mathcal{NP}$ by measuring the difference in dissipative computation due to the difference in degrees of freedom. Moreover, since the computation does not advance by non-dissipative (reversible) transitions, these do not affect the measure.

To maintain a connection to practicalities, it is worth noting that tractable problems are often idealizations of intractable natural processes. For example, when determining the shortest path for a long-haul trucker to take through a network of cities to the destination, it is implicitly assumed that when the computed optimal path is actually taken, the traffic itself would not congest the current and cause a need for rerouting and finding a new, best possible route under the changing circumstances.

The state space of a finite energy system is represented by elements $\phi$ of the set $\Phi$ (45). Transformations from a state to another are represented by elements $\lambda$, referred to as process generators of the set $\Lambda$. The computation is a series of transformations along a piecewise continuous path $s(\lambda, \phi)$ in the state space. According to the 2$^{nd}$ law the paths of energy dispersal that span the affine manifold $\mathcal{M}$, are shortening until the free energy minimum state has been attained. Then the state space has contracted during the transformation process to the accepting state.

**Definition 8.1** A system is a pair ($\Phi$, $\Lambda$), with $\Phi$ a set whose elements $\phi$ are called states and $\Lambda$ a set whose elements $\lambda$ are called process generators, together with two functions. The function $\lambda \mapsto \mathcal{S}$ assigns to each $\lambda$ a transformation $\mathcal{S}$, whose domain $\mathcal{D}(\lambda)$ and range $\mathcal{R}(\lambda)$ are non-empty subsets of $\Phi$ such that for each $\phi$ in $\Phi$ the condition of accessibility holds

$$(i) \ \Lambda\phi := \{\mathcal{S}\phi : \lambda \in \Lambda, \phi \in \mathcal{D}(\lambda)\} = \Phi \qquad (8.1a)$$

where $\Lambda\phi$ is the entire set of states accessible from $\phi$, with the assertion that, for every state $\phi$, $\Lambda\phi$ equals the entire state space $\Phi$. Furthermore, the function $(\lambda´, \lambda´´) \mapsto \lambda´´\lambda´$ assigns to each pair $(\lambda´, \lambda´´)$ the (extended) process generator $\lambda´´\lambda´$ for the successive application of $\lambda´´$ and $\lambda´$ with the property:

$$\begin{aligned}(ii) \ &\text{if } \mathcal{D}(\lambda´´) \cap \mathcal{R}(\lambda´) \neq \varnothing, \text{ then} \\ &\mathcal{D}(\lambda´´\lambda´) = \mathcal{S}_{\lambda´}^{-1}(\mathcal{D}(\lambda´´)) \\ &\text{and, for each } \phi \text{ in } \mathcal{D}(\lambda´´\lambda´) \text{ there holds} \\ &\mathcal{S}_{\lambda´´\lambda´}\phi = \mathcal{S}_{\lambda´´}\mathcal{S}_{\lambda´}\phi, \forall \phi \in \mathcal{D}(\lambda´´\lambda´) \\ &\text{when for any other } \lambda^* \\ &\mathcal{D}(\lambda´) \cap \mathcal{D}(\lambda^*) = \varnothing. \end{aligned} \qquad (8.1b)$$

The extended process generators $\lambda´´\lambda´$ formalize the successive transformations with less than three degrees of freedom. When the transformation $\mathcal{S}_{\lambda´}$ is emissive, its inverse $\mathcal{S}_{\lambda´}^{-1}$ is absorptive.

**Definition 8.2** A process of ($\Phi$, $\Lambda$) is a pair ($\lambda$, $\phi$) such that $\phi \in \mathcal{D}(\lambda)$. The process generators transform the system from an initial state via intermediate states to the final state. The set of all processes of ($\Phi$, $\Lambda$) is

$$\Lambda \diamond \Phi = \{(\lambda, \phi) : \lambda \in \Lambda, \phi \in \mathcal{D}(\lambda)\}. \qquad (8.2)$$

According to definitions 8.1 and 8.2 the states and process generators are interdependent (Fig. 5) so that:

(i) When the system has transformed from the state $\phi$ to the state $\mathcal{S}_\lambda\phi$, the process generator $\lambda$ has vanished.

(ii) When the system has transformed from $\phi$ to $\mathcal{S}_\lambda\phi$, the system is no longer at $\phi$ available for another transformation $\mathcal{S}_{\lambda^*}$ by another process generator $\lambda^*$ to $\mathcal{S}_{\lambda^*}\phi$.

(iii) When the system has transformed from the initial state $\phi$ to an intermediate state $\mathcal{S}_{\lambda´}\phi$ and subsequently from $\mathcal{S}_{\lambda´}\phi$ to $\mathcal{S}_{\lambda´´}\mathcal{S}_{\lambda´}\phi$, the final state $\mathcal{S}_{\lambda´´}\mathcal{S}_{\lambda´}\phi$ is identical to the state resulting from the extended transformation from $\phi$ to $\mathcal{S}_{\lambda´´\lambda´}\phi$, only when $\mathcal{S}_{\lambda´}\phi$ is not a domain $\mathcal{D}(\lambda^*)$ of any other transformation $\mathcal{S}_{\lambda^*}$.

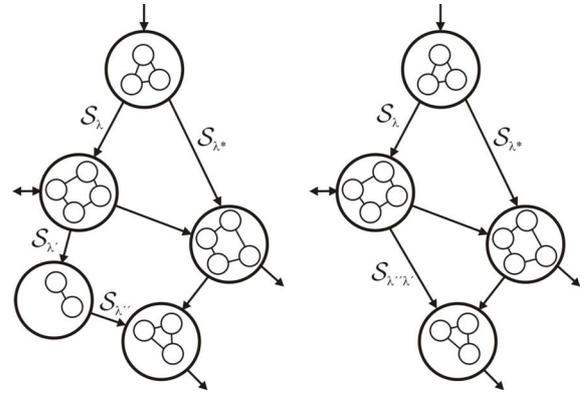

Figure 5. (Left) The system evolves, according to the definitions 8.1 and 8.2, from an initial state (top) to other states by a sequence of transformations $\mathcal{S}$ (arrows) that are directional, *i.e.*, dissipative due to the distinct domains $\mathcal{D}$ and ranges $\mathcal{R}$ for distinct elements $\lambda$ of process generators. (Right) The successive transformations $\mathcal{S}_{\lambda´}$ and $\mathcal{S}_{\lambda´´}$ can be reduced to $\mathcal{S}_{\lambda´´\lambda´}$ only when the intermediate state cannot be transformed by any other process $\lambda^*$.

**Definition 8.3** (45) Let $t > 0$ and let $\lambda_t$: $[0, t) \mapsto \mathbb{R}$, be piecewise continuous, and define $\mathcal{D}(\lambda_t)$ to be the set of states $\phi = (N, G) \in \Phi$ such that the differential equation

$$\left(\frac{dN(\tau)}{dt}, \frac{dG(\tau)}{dt}\right) = \lambda_t(\tau) \qquad (8.3)$$

has a solution $\tau \mapsto (N(\tau), G(\tau))$ that satisfies the initial condition $(N(0), G(0)) = \phi$ and follows the trajectory $\{(N(\tau), G(\tau)) | \tau \in [0, t]\}$ which is entirely in $\Phi$. In other words, $\phi \in \mathcal{D}(\lambda_t)$ if and only if $\phi + \int_0^\tau \lambda_t(\xi)d\xi$ is in $\Phi$ for every $\tau \in [0, t]$.

When Eq. 8.3 is compared with Eq. 4.1, $\lambda_t$ is understood in the continuum limit to generate a transformation from the initial density $\phi = (N(0), G(0))$ (*cf.* the definition of energy density in Sec. 3) to a succeeding density $\phi_\tau = (N(\tau), G(\tau))$ during a step $\tau \in [0, t]$ via the flow $v = dN/dt$ that drains the free energy.

**Definition 8.4** (45) Define $\Lambda$ to be the set of functions $\lambda_t$ for which $\mathcal{D}(\lambda_t) \neq \emptyset$. For each $\lambda_t \in \Lambda$, define $\mathcal{S}_{\lambda t}\phi: \mathcal{D}(\lambda_t) \mapsto \Phi$ by the formula

$$\mathcal{S}_{\pi_t}\phi = \phi + \int_0^t \lambda_t(\xi)d\xi. \qquad (8.4)$$

If $s(\lambda_t, \phi)$ denotes the path determined by $\tau \mapsto \phi + \int_0^\tau \lambda_t(\xi)d\xi$ $\tau \in [0, t]$, then $\mathcal{S}_{\lambda t}\phi$ is taken to be the final point of $s(\lambda_t, \phi)$. Moreover $\phi \in \mathcal{D}(\lambda_t) \Leftrightarrow s(\lambda_t, \phi) \subset \Phi$.

The step of evolution along the oriented and piecewise smooth curve from $\phi$ to $\mathcal{S}_{\lambda t}\phi$ is the path $s(\lambda_t, \phi) \subset \Phi$ determined by the formal integration from 0 to $\tau$ (Eq. 8.4). In the general case of dissipative transformations with degrees of freedom ($n \geq 3$) the integration is not closed. An open system is spiraling along an open trajectory either by loosing quanta to or acquiring them from its surroundings. Consequently the state space $\phi \in \mathcal{D}(\lambda_t)$ is contracting by successive applications of $\lambda_t'$ and $\lambda_t''$ that diminish the free energy almost everywhere such that $\mathcal{R}(\lambda_t') \subseteq \mathcal{D}(\lambda_t')$. The dissipation ceases first at the free energy minimum state where the orbits are closed and the domain and range are indistinguishable for any process.

**Definition 8.5** (46) After a series of successive applications of $\lambda_t''$ and $\lambda_t'$ the evolving system arrives at the free energy minimum. Then the open system is in a dynamic state defined as the $\varepsilon$-steady state by a fixed non-zero set $\varepsilon = \{\varepsilon_S\}$ such that during $\tau$ if and only if, for all $\mathcal{S} \in \Phi$, there exists $\zeta_S \in \mathbb{R}$ such that for all $\tau \in \Phi$, it follows

$$\left|\langle \mathcal{S} \rangle_\tau - \zeta_S\right| \leq \varepsilon_S. \qquad (8.5)$$

At the $\varepsilon$-steady state there is no net flux over the period of integration $\tau \in [0, t]$. Thus the probability $P$ may fluctuate due to sporadic influx and efflux but its absolute value may not exceed $\varepsilon_S$ so that the system continues to reside within $\varepsilon$. The set value $\varepsilon_S$ defines the acceptable state of computation, otherwise in the continuum limit $\varepsilon \to 0$ the state space would contract indefinitely. In practice the state space sampling by brute-force algorithms or simulated annealing methods is limited by $\varepsilon_S$, *e.g.*, according to the available computational resources.

**Definition 8.6** (73) A family $\Sigma$ of subsets of the state space $\Phi$ is an algebra, if it has the following properties:

(i) $\Phi \in \Sigma$,
(ii) $\Phi_0 \in \Sigma \Rightarrow \Phi_0^c \in \Sigma$
(iii) $\{\Phi_i\}_{i \in [1,k]} \subset \Sigma \Rightarrow \bigcup_{i=1}^k \Phi_i \in \Sigma$ (8.6)

from these it follows

(i) $\emptyset \in \Sigma$,
(ii) the algebra $\Sigma$ is closed under countable intersections and subtraction of sets, and
(iii) if $k \equiv \infty$ then $\Sigma$ is said to be a sigma-algebra.

**Definition 8.7** (73) A function $\mu_C: \Sigma \mapsto [0, \infty)$ is a measure if it is additive for any countable subfamily $\{\Phi_i, i \in [1,n]\} \subseteq \Sigma$, consisting of mutually disjoints sets, such that

$$\mu_C\left(\bigcup_{i=1}^n \Phi_i\right) = \sum_{i=1}^n \mu_C(\Phi_i)$$

It follows: (8.7)

(i) $\mu_C(\emptyset) = 0$,
(ii) if $\Phi_\alpha, \Phi_\beta \in \Sigma$ and $\Phi_\alpha \subset \Phi_\beta \Rightarrow \mu_C(\Phi_\alpha) \leq \mu_C(\Phi_\beta)$,
(iii) and if $\Phi_\alpha, \Phi_\beta \in \Sigma$ and $\Phi_1 \subset \Phi_2 \subset \cdots \subset \Phi_n$ and $\{\Phi_i, i \in [1,n]\} \in \Sigma \Rightarrow \mu_C\left(\bigcup_{i=1}^n \Phi_i\right) = \sup_i \mu_C(\Phi_i)$.

Moreover, if $\Sigma$ is a sigma-algebra and $n \equiv \{\infty\}$, then $\mu_C$ is said sigma-additive. The triple $(\Phi, \Sigma, \mu_C)$ is a measure space.

**Definition 8.8** (45) An energy density manifold is a set $\mathcal{M}$ whose elements $\phi$ are called energy densities together with a set $\Sigma$ of functions $\mu_i: \mathcal{M} \mapsto \mathbb{R}$ called energy scale, satisfying:

(i) The range of $\mu$ is an open interval for each $\mu_i \in \Sigma$,
(ii) for every $\phi_A, \phi_B \in \mathcal{M}$ and $\mu \in \Sigma$, (8.8a)
$\mu_A(\phi_A) = \mu_B(\phi_B) \Rightarrow \phi_A = \phi_B$
(iii) for every $\mu_A, \mu_B \in \Sigma$, $\theta \mapsto \mu_B(\mu_A^{-1}(\theta))$
is a continuous, strictly increasing function.

(i) asserts that each energy scale takes on all values in an open interval in $\mathbb{R}$, while (ii) guarantees that each such scale establishes a one-to-one correspondence between energy levels and real numbers in its range. By means of (iii) the set $\Sigma$ determines an order relation $\prec$ on $\mathcal{M}$ written as:

$$\phi_A \prec \phi_B \Leftrightarrow \text{there exists } \mu_i \in \Sigma \qquad (8.8b)$$
$$\text{such that } \mu_A(\phi_A) < \mu_B(\phi_B).$$

Physically speaking the energy densities are in relation to each other on the energy scale given in the units of $\theta = k_B T$.

**Definition 8.9** Entropy is defined as

$$S = \sum_{j=1} k_B \ln P_j = \sum_{j=1} k_B N_j \left(1 - \sum_{j,k} \frac{\Delta V_{jk}}{k_B T}\right) \quad (8.9)$$

where the absolute temperature $T > 0$ and the Boltzmann's constant $k_B > 0$ in accordance with the equation 3.3.

**Definition 8.10** The change in occupancy $N_j$ is defined proportional to the free energy

$$\Delta N_j = -\sum_k \sigma_{jk} \frac{\Delta V_{jk}}{k_B T} \quad (8.10)$$

in accordance with the equation 3.4.

**Theorem 8.11** *The principle of increasing entropy.* The condition of stationary state for the open system is that its entropy reaches the maximum.

*Proof.* From the definitions 8.9 and 8.10 and $\Delta \mu_{jk}(\Delta N_j)/k_B T = \Delta N_j/N_j$, it follows that

$$\Delta S = k_B L = -k_B \sum_{j=1} \Delta N_j \sum_k \frac{\Delta V_{jk}}{k_B T} = k_B \sum_{j,k} \sigma_{jk}^{-1} (\Delta N_j)^2$$
$$= k_B \sum_{j,k} \sigma_{jk} \left(\frac{\Delta V_{jk}}{k_B T}\right)^2 \geq 0 \quad (8.11)$$

because the squares are non-negative, the conductance $\sigma_{jk} > 0$ and its inverse, i.e., resistance, $\sigma_{jk}^{-1} = m_{jk}/k_B T > 0$ and $k_B > 0$.

The proof is in agreement with $\Delta S = k_B \Delta \ln P = k_B L \geq 0$ given by Eq. 4.1. The principle of increasing entropy has been proven alternatively by variations $\delta$ using the principle of least action $\delta A := \delta_0 \int^t \mathcal{L} dt = -\delta_0 \int^t TS dt \leq 0$ (46) where the Lagrangian $\mathcal{L}$ integrand (kinetic energy) defined by the Gouy-Stodola theorem, is necessarily positive.

**Theorem 8.12** *The state space $\Phi$ contracts in dissipative transformations.*

*Proof.* As a consequence of the definitions 8.10 and 8.11 it follows that

$$\Delta(\Delta S) = k_B \Delta L = 2 \sum_{j=1} \frac{\Delta N_j}{N_j} \sum_k \sigma_{jk} \frac{\Delta V_{jk}}{T} \quad (8.12)$$
$$= -2 k_B \sum_{j=1} \frac{(\Delta N_j)^2}{N_j} \leq 0$$

because the squares are non-negative, the occupancies $N_j > 0$ for non-zero densities-in-energy, the conductance $\sigma_{jk} \geq 0$, $T > 0$ and $k_B > 0$.

When entropy $S$ is increasing, the state space accessible by the process generator $L$ is decreasing. In the continuum limit the theorem for contraction has been proven earlier (46). In practice the contraction of the state space by a finite automaton is limited to a fixed non-zero set $\varepsilon = \{\varepsilon_S\}$. Then any member in $\varepsilon$ is qualified as solution.

**Definition 8.13** The definition for the class $\mathcal{P}$ state space measure $\mu_\mathcal{P}$ follows from the definitions 8.7 and 8.9

$$\mu_\mathcal{P} = \ln P_\mathcal{P} = \sum_{j=1}^n N_j \left(1 - \sum_{k \neq j}^n \frac{\Delta \mu_{jk}}{k_B T}\right)$$
$$+ \sum_{j=1}^n N_j \sum_{k=j\pm 1}^n \frac{\Delta Q_{jk}}{k_B T}. \quad (8.13)$$

The non-dissipative (reversible) and dissipative (irreversible) components have been denoted separately. In fact, the indexing $k \neq j$ is redundant because for the indistinguishable sets $k = j$ there is no difference, per definition $\Delta \mu_{jj} = 0$. The conserved term $\Sigma_j N_j(1-\Sigma_{k\neq j}\Delta\mu_{jk})$ is invariant according to Noether's theorem (24). The non-zero dissipative term $\Sigma_j N_j \Sigma_{k=j\pm 1} \Delta Q_{jk}$ defines class $\mathcal{P}$ to contain at least one irreversible deterministic decision with two degrees of freedom ($n = 2$).

**Definition 8.14** The definition for the class $\mathcal{NP}$ state space measure $\mu_{\mathcal{NP}}$ follows from the definitions 8.7 and 8.9

$$\mu_{\mathcal{NP}} = \ln P_{\mathcal{NP}} = \sum_{j=1}^n N_j \left(1 - \sum_{k \neq j}^n \frac{\Delta \mu_{jk}}{k_B T}\right)$$
$$+ \sum_{j=1}^n N_j \sum_{k=j\pm 1}^n \frac{\Delta Q_{jk}}{k_B T} + \sum_{j=1}^n N_j \sum_{k \neq j\pm 1}^n \frac{\Delta Q_{jk}}{k_B T}. \quad (8.14)$$

The conserved components have been denoted separately from the dissipative components that have been decomposed further to those with two degrees of freedom using the indexing notation $k = j \pm 1$ as well as to those with three or more degrees of freedom using the indexing notation $k \neq j \pm 1$. The conserved and dissipative components with only two degrees of freedom are the same as those in definition 8.13. The non-zero dissipative term $\Sigma_j N_j \Sigma_{k \neq j \pm 1} \Delta Q_{jk}$ defines class $\mathcal{NP}$ to contain at least one irreversible decision between at least two choices, i.e., with the three or more degrees of freedom.

**Definition 8.15** The $\mathcal{NP}$-complete problem contains only dissipative processes with three or more degrees of freedom, *i.e.*, $\Sigma_j N_j \Sigma_{k \neq j \pm 1} \Delta Q_{jk} > 0$ and none with two degrees of freedom $\Sigma_j N_j \Sigma_{k = j \pm 1} \Delta Q_{jk} = 0$.

**Theorem 8.16** $\mathcal{P} \subset \mathcal{NP}$.

*Proof*. It follows from the definitions 8.13 and 8.14 that the state space set of class $\mathcal{NP}$ is larger than class $\mathcal{P}$ measured by the difference

$$\mu_{\mathcal{NP}-\mathcal{P}} = \mu_{\mathcal{NP}} - \mu_{\mathcal{P}} = \sum_{j=1}^n N_j \sum_{k \neq j \pm 1}^n \frac{\Delta Q_{jk}}{k_B T} > 0. \quad (8.16)$$

If and only if $\Delta Q_{jk} = 0$ for all $k \neq j \pm 1$, the measure $\mu_{\mathcal{NP}-\mathcal{P}}(\varnothing) = 0$ but this is a contradiction with definition 8.14 that class $\mathcal{NP}$ contains at least one irreversible decision with three or more degrees of freedom, *i.e.*, $\Sigma_j N_j \Sigma_{k \neq j \pm 1} \Delta Q_{jk} > 0$. Thus class $\mathcal{P}$ is a proper subset of class $\mathcal{NP}$.

The difference between the classes can also be measured by $P_{\mathcal{NP}} \ln(P_{\mathcal{NP}}/P_{\mathcal{P}}) > 0$ in accordance with the non-commutative measure known as Gibb's inequality or Kullback–Leibler divergence that gives the difference between two probability distributions.

The class $\mathcal{NP}$ problem can be reduced to the class $\mathcal{NP}$-complete problem by removing the deterministic steps denoted by $k = j \pm 1$, *i.e.*, by polynomial time reduction (23,74). In graphical terms the reduction of the $\mathcal{NP}$ problem to the $\mathcal{NP}$-complete problem involves removal of nodes with less than three degrees of freedom (Fig. 6). In geometric terms the non-Euclidean landscape is reduced to a manifold covered by non-equivalent triangles each having a local Lorentzian metric.

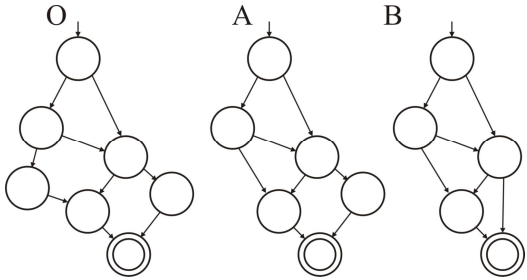

Figure 6. The network representing the class $\mathcal{NP}$ problem (O) is reduced (O → A → B) to the network representing the class $\mathcal{NP}$-complete problem by removing nodes along deterministic dissipative paths to yield a network of triangles.

In summary the computational complexity classes are related to each other as $\mathcal{P} \subset \mathcal{NP}$-C $\subseteq \mathcal{NP}$ (Fig. 7).

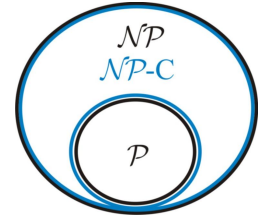

Figure 7. Venn diagram for the computational complexity classes $\mathcal{P}$, $\mathcal{NP}$-complete and $\mathcal{NP}$ based on the thermodynamic analysis of computation. The class $\mathcal{P}$ problems can be computed by dissipative processes that have less than three degrees of freedom whereas the class $\mathcal{NP}$ problem computation involves in addition dissipative processes with three or more degrees of freedom. The class $\mathcal{NP}$-complete problem computation contains only dissipative processes with three or more degrees of freedom.

## 9. Discussion

At first sight it may appear strange for some that the distinction between the computational complexity class $\mathcal{P}$ and $\mathcal{NP}$ was made on the basis of the natural law because both classes contain many abstract problems without apparent physical connection. However, the view is not new (75,76,77,78). The adopted approach on the classification of computational complexity is motivated because the practical computation is a thermodynamic process hence inevitably subject to the 2$^{nd}$ law of thermodynamics. Of course, some may still argue that the distinction between tractable and intractable problems ought to be proven without any reference to physics. Indeed, the physical portrayal can be taken merely as a formal notation to express that the computation is a series of time-ordered operations that are intractable when there are three or more degrees of freedom among interdependent operations. Also non-commutative operations and non-abelian groups formalize time series (79,80). The essential ingredient is that decisions affect set of future decisions, *i.e.*, the driving forces of computation depend on the process itself. The formulation by the 2$^{nd}$ law of thermodynamics is a natural expression because the free energy and the flow of energy are naturally interdependent.

The natural law may well be the invaluable ingredient to rationalize the distinction between the computational complexity classes $\mathcal{P}$ and $\mathcal{NP}$. It serves not only to prove that $\mathcal{P} \subset \mathcal{NP}$ but to account for the computational course itself. For both classes of problems the natural process of computation is directing toward increasingly more probable states. When there are three or more degrees of freedom, decisions influence the choice of future decisions and the computation is intractable. The set of conceivable states generated at the branching points can be enormous, similar to a causal Bayesian network (81). Finally, when the maximum entropy state has been attained, it can be validated independent of the path as the free energy minimum stationary state. The corresponding solution is verifiably independent of the computational history in polynomial time.

Furthermore, the crossing from class $\mathcal{P}$ to $\mathcal{NP}$ is found precisely where $n$-SAT, $n$-coloring, $n$-clique problems and maximizing the shortest path with interdicts become intractable, i.e., when the degrees of freedom $n \geq 3$. The efficient reduction of $\mathcal{NP}$ problems to $\mathcal{NP}$-complete problems is also understood as operations that remove the deterministic dissipative steps and eventual redundant reversible paths. Besides, when the problem is beyond class $\mathcal{NP}$, the natural process does not terminate at the accepting state with emission. For example, the halting problem belongs to the class $\mathcal{NP}$-hard. Importantly, the natural law relates computational time directly to the flow of energy, i.e., to the amount of dissipation (10). Thus the 2$^{nd}$ law implies that non-dissipative processing protocols are deemed futile (82).

The practical value of computational complexity classification by the natural law of the maximal energy dispersal is that no deterministic algorithm can be found that would complete the class $\mathcal{NP}$ problems in polynomial time. The conclusion is anticipated (83), nonetheless, its premises imply that there is no all-purpose algorithm to trace the maximal flow paths through non-invariant landscapes. Presumably the most general and efficient algorithms balance execution between exploration of the landscape and progression down along the steep gradients in time. Perhaps most importantly, the universal law provides us with a holistic understanding of the phenomena themselves to formulate computational tasks in the most meaningful way.

**Acknowledgments.** I am grateful to Mahesh Karnani, Heikki Suhonen and Alessio Zibellini for valuable corrections and instructive comments.